\documentclass[twocolumn,showkeys,aps,prb,showpacs]{revtex4-1}
\UseRawInputEncoding
\usepackage{graphicx}
\usepackage[CJKbookmarks,dvipdfm,colorlinks,linkcolor=red,citecolor=blue]{hyperref}

\begin{document}

\title{A possible electronic state quasi-half-valley-metal in $\mathrm{VGe_2P_4}$ monolayer}

\author{San-Dong Guo$^{1}$,  Yu-Ling Tao$^{1}$, Zhuo-Yan  Zhao $^{1}$, Bing Wang$^{2}$,  Guangzhao Wang$^{3}$ and  Xiaotian Wang$^{4}$}
\affiliation{$^1$School of Electronic Engineering, Xi'an University of Posts and Telecommunications, Xi'an 710121, China}
\affiliation{$^2$Institute for Computational Materials Science, School of Physics and Electronics,Henan University, 475004, Kaifeng, China}
\affiliation{$^3$Key Laboratory of Extraordinary Bond Engineering and Advanced Materials Technology of Chongqing, School of Electronic Information Engineering, Yangtze Normal University, Chongqing 408100, China}
\affiliation{$^4$Institute for Superconducting and Electronic Materials, University of Wollongong, Wollongong 2500, Australia} 
\begin{abstract}
One of the key problems in valleytronics is to realize valley polarization. Ferrovalley (FV) semiconductor and half-valley-metal (HVM) have been proposed,
which possess intrinsic spontaneous valley polarization. Here, we propose the concept of quasi-half-valley-metal (QHVM), including electron and hole carriers with only a type of carriers being valley polarized. The QHVM may realize separation function of electron and hole.   A concrete example of $\mathrm{VGe_2P_4}$ monolayer is used to  illustrate our  proposal through the first-principle calculations. To better realize QHVM, the electric field is applied to tune related valley properties of  $\mathrm{VGe_2P_4}$.
Within considered  electric field range, $\mathrm{VGe_2P_4}$ is always ferromagnetic (FM) ground state, which possesses  out-of-plane magnetization by calculating magnetic anisotropy energy (MAE) including magnetic shape anisotropy (MSA) and magnetocrystalline anisotropy (MCA) energies.
These out-of-plane FM properties guarantee  intrinsic spontaneous valley polarization in $\mathrm{VGe_2P_4}$.
  Within a certain range of electric field, the  QHVM can be maintained, and the related polarization properties can be effectively tuned.
 Our works  pave the way toward   two-dimensional (2D) functional  materials design of valleytronics.

\end{abstract}
\keywords{Valley, Electric field, Magnetic anisotropy  ~~~~~~~~~~~~~Email:sandongyuwang@163.com}

\maketitle

\section{Introduction}
In analogy to charge/spin of electronics/spintronics, the valley has been recognized as an extra degree of
freedom  for carriers, widely
known as valleytronics\cite{b0}. The valley is characterized by a local energy extreme in
the conduction band or valence band.  In a material, two or more degenerate but inequivalent valley states  should exist to encode, store and process information\cite{b1,b2}.
For nonmagnetic valleytronic materials, 2H-transition-metal dichalcogenides (TMD) have
attracted extensive attention\cite{q8-1,q8-2,q8-3,q9-1,q9-2,q9-3,q9-4}, because they have a pair of
degenerate but inequivalent valleys  and the large intervalley
distance. However, these 2D materials lack spontaneous valley polarization.
Although various strategies, such as  optical pumping, magnetic field, magnetic
substrates and  magnetic doping\cite{q8-1,q8-2,q8-3,q9-1,q9-2,q9-3,q9-4}, have
been executed  to artificially
induce valley polarization, these methods destroy
the  intrinsic  energy band structures and crystal structures.

\begin{figure}
  \includegraphics[width=8cm]{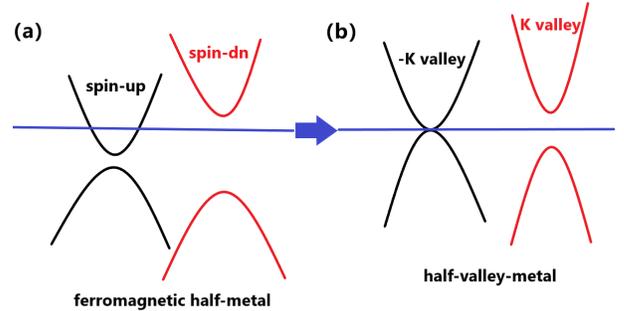}
  \caption{(Color online) Schematic diagram of the analogy between ferromagnetic half-metal (a) and half-valley-metal (b).
   The spin-up/spin-dn is equivalent to -K/K valley. In (a), one spin channel is conducting, whereas the other is insulating. In (b),  one valley channel is metal, in which conduction electrons are intrinsically
 100\% valley polarized, and the other is insulator.}\label{sy0}
\end{figure}

Fortunately,  FV semiconductor with spontaneous
spin and valley polarizations has been proposed\cite{q10}, which breaks time-reversal and
space-inversion symmetry with perfect coupling of the valley with charge and spin.
The  FV  semiconductors have been predicted in many 2D materials\cite{q10,q11,q12,q13,q13-1,q19,gsd,q14,q14-0,q14-0-1,q14-0-2,q14-0-3,q14-1,q17,q18,q14-2,q14-3,q14-4,q15,q16}, which  can overcome these shortcomings of the extrinsic
valley polarization materials. Recently, in analogy to half-metals in spintronics (see \autoref{sy0}),  the concept of HVM has been proposed\cite{q10-1}.
The conduction electrons of HVM are intrinsically 100\% valley polarized and  100\% spin polarized even when including spin-orbit coupling (SOC).
The electron correlation effect or strain can induce the FV semiconductor to HVM transition in some special 2D materials\cite{q14-0,q10-1,q10-2,a6,a7}, which is generally related to topological phase transitions. However, the HVM is just at one point, not a region of electron correlation strength or strain, which may be difficult to achieve in experiment.

\begin{figure*}
  \includegraphics[width=16cm]{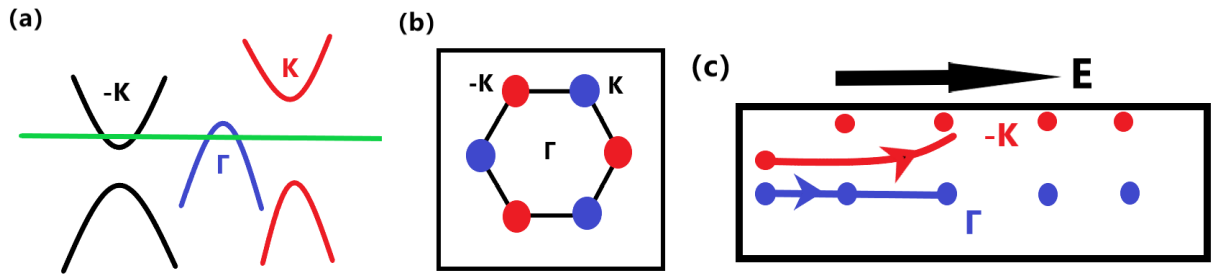}
  \caption{(Color online) Schematic diagram of quasi-half-valley-metal: (a) the energy band structures, and the Fermi level slightly touches the CBM   and VBM. (b): the distribution of Berry curvature in 2D BZ, and Berry curvature only occurs around -K and K valleys with opposite
signs and unequal magnitudes. (c):  under an in-plane longitudinal electric field $E$,
    the  electronic carriers  of -K valley  turn towards one edge of the sample, and  the hole carriers of $\Gamma$ valley move in a straight line.}\label{sy}
\end{figure*}

In this work, we   propose a  concept of QHVM, where electron and hole carriers simultaneously exist with only a type of carrier being valley polarized (see \autoref{sy}). In QHVM, the Fermi level slightly touches the conduction band minima (CBM)   and valence band maxima (VBM), and the  Berry curvature in 2D Brillouin zone (BZ) only occurs around -K and K valleys with opposite
signs and unequal magnitudes. Under an in-plane longitudinal electric field $E$, the nonzero Berry curvature $\Omega (k)$ makes
the  carriers of -K valley obtain the general group velocity $v_{\parallel}$ and the
anomalous transverse velocity $v_{\bot}$\cite{xd,qqq}:
\begin{equation}\label{d-d-1}
v=v_{\parallel}+v_{\bot}=\frac{1}{\hbar}\nabla_k\varepsilon(k)-\frac{e}{\hbar}E\times\Omega(k)
 \end{equation}
where $v_{\parallel}$ ($v_{\bot}$) is along the electric field direction  (perpendicular to the electric field and out-of-plane directions).
However,  carriers of $\Gamma$ valley only obtain the general group velocity $v_{\parallel}$. The QHVM may be used to separate electron and hole carriers.

Recently, 2D $\mathrm{MA_2Z_4}$
family  with a septuple-atomic-layer structure  has been established\cite{a11,a12} with   diverse
properties from common semiconductor to FV semiconductor to topological insulator to Ising superconductor, and $\mathrm{MoSi_2N_4}$ and $\mathrm{WSi_2N_4}$ of them
  have been
achieved experimentally  by the chemical vapor deposition
method. Taking the  $\mathrm{VGe_2P_4}$ monolayer  as an example, the QHVM can be realized by electric filed.
Within considered  electric field range, the out-of-plane FM properties guarantee  intrinsic spontaneous valley polarization in $\mathrm{VGe_2P_4}$.
 The  QHVM can be maintained in a certain range of electric field, and the related polarization properties can be effectively tuned.
Our findings may be extended to other FV semiconductors, and the QHVM can be achieved by electric field tuning.

The rest of the paper is organized as follows. In the next
section, we shall give our computational details and methods.
 In  the next few sections,  we shall present structure and stabilities, magnetic anisotropy (MA) and electronic structures  and  electric field effects on  physical properties  of  $\mathrm{VGe_2P_4}$ monolayer. Finally, we shall give our discussion and conclusion.

\begin{figure*}
  \includegraphics[width=16cm]{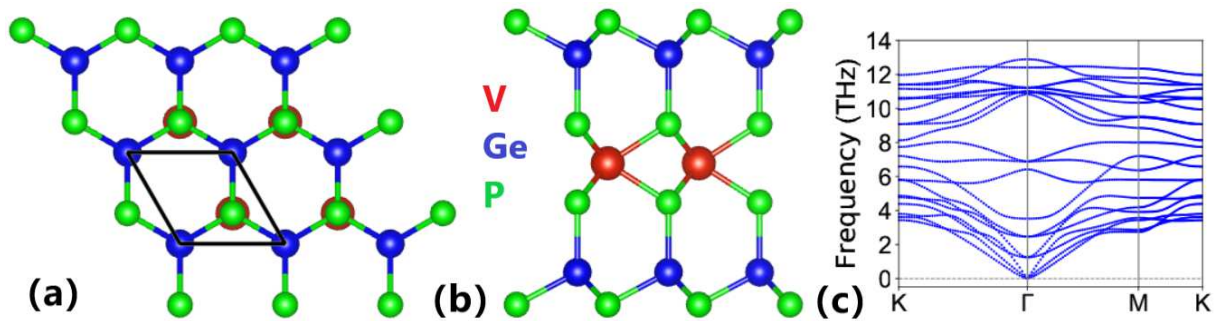}
  \caption{(Color online) For $\mathrm{VGe_2P_4}$ monolayer,  (a): top view and (b): side view of  crystal structure,  and the  primitive  cell is
   shown by black lines in (a). (c): the phonon dispersions  with FM order using GGA+$U$ ($U$= 3 eV). }\label{st}
\end{figure*}
\begin{figure}
  \includegraphics[width=8cm]{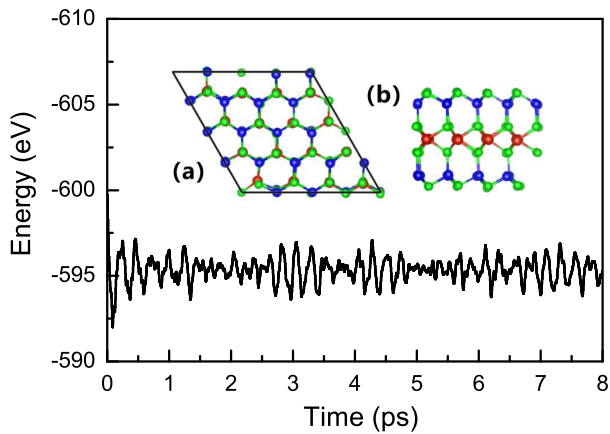}
  \caption{(Color online) For $\mathrm{VGe_2P_4}$ monolayer, the variation of free energy during the 8 ps AIMD simulation  using GGA+$U$ ($U$= 3 eV). Insets show the
 final structures (top view (a) and side view (b))  after 8 ps at 300 K. }\label{md}
\end{figure}

\begin{figure*}
  \includegraphics[width=15cm]{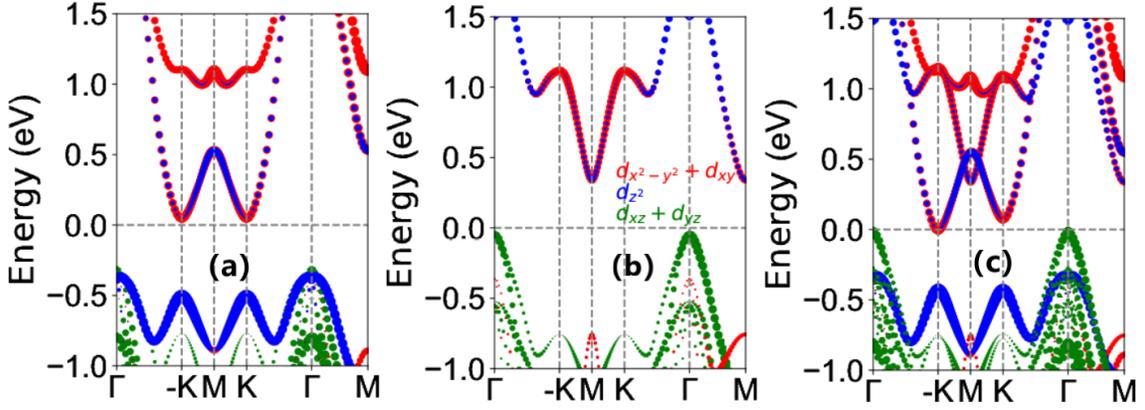}
  \caption{(Color online) For $\mathrm{VGe_2P_4}$ monolayer,  the V-$d_{x^2-y^2}$+$d_{xy}$, $d_{z^2}$ and $d_{xz}$+$d_{yz}$-orbital characters energy band structures  without SOC (a, b) and  with SOC (c)  using GGA+$U$ ($U$= 3 eV). The (a) and (b)  represent the band structure in the spin-up and spin-dn directions. }\label{band0}
\end{figure*}

\begin{figure}
  \includegraphics[width=8cm]{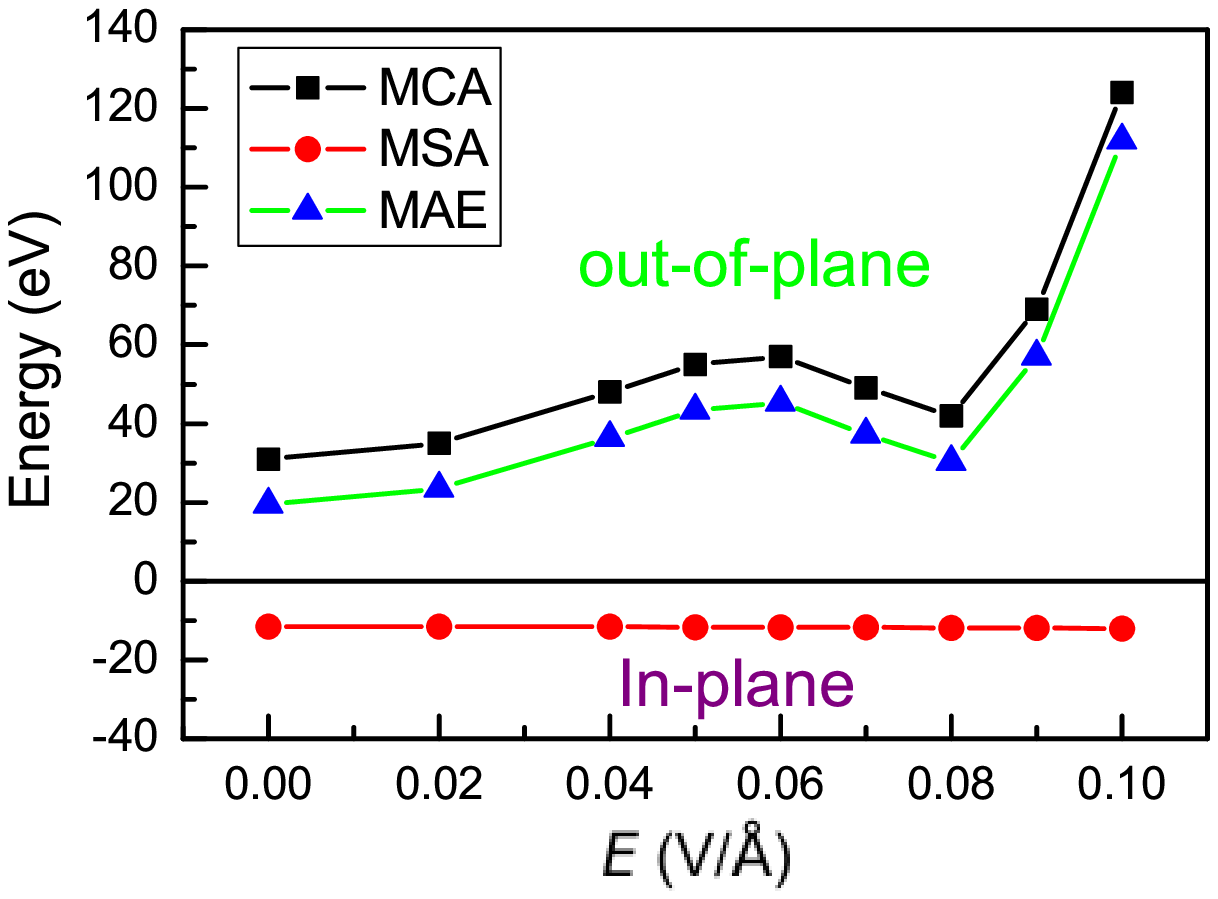}
  \caption{(Color online) For $\mathrm{VGe_2P_4}$ monolayer, the  MCA energy, MSA energy  and MAE  as a function of $E$ using GGA+$U$ ($U$= 3 eV). }\label{mae}
\end{figure}

\section{Computational detail}
Within density-functional
theory  (DFT)\cite{1},  the spin-polarized  first-principles calculations are performed  by employing the projected
augmented wave method,  as implemented in Vienna ab initio simulation package (VASP)\cite{pv1,pv2,pv3}.
We use the generalized gradient approximation of Perdew-Burke-Ernzerhof (PBE-GGA)\cite{pbe} as exchange-correlation functional.
The on-site Coulomb correlation of V  atoms is considered by the GGA+$U$   method   in terms of  the on-site Coulomb
interaction of $U$$=$ 3.0 eV, as widely used in V-based $\mathrm{MA_2Z_4}$ family\cite{q14,q14-0-1,q14-0-2,q14-0-3}.
The GGA+$U$ adopts  the rotationally invariant approach proposed by Dudarev et al\cite{u}, in which  only the effective
$U$ ($U_{eff}$) based on the difference between the on-site Coulomb interaction
parameter  and exchange parameters  is meaningful.

To attain accurate results, we use the energy cut-off of 500 eV,  total energy  convergence criterion of  $10^{-8}$ eV and  force
convergence criteria of less than 0.0001 $\mathrm{eV.{\AA}^{-1}}$ on each atom.
A vacuum space of more than 20 $\mathrm{{\AA}}$ is used to avoid the interactions
between the neighboring slabs.
  A $\Gamma$-centered 16 $\times$16$\times$1 k-point meshs  in  the BZ  are used for these calculations of structure optimization, electronic structures and elastic constants, and a 9$\times$16$\times$1 Monkhorst-Pack k-point meshs for calculating FM/antiferromagnetic (AFM)  energy  with rectangle supercell, as shown in FIG.1 of electronic supplementary information (ESI) along with the first BZ.
 The SOC effect is explicitly included to investigate MCA and  electronic  properties of  $\mathrm{VGe_2P_4}$ monolayer.

 Through the direct supercell method,  the interatomic force constants (IFCs) using GGA+$U$ are calculated
  with the 5$\times$5$\times$1 supercell. Based on calculated  harmonic IFCs, the
phonon dispersions are obtained by using Phonopy code\cite{pv5}.
The elastic stiffness tensor  $C_{ij}$   are carried out by using strain-stress relationship (SSR),  and the 2D elastic coefficients $C^{2D}_{ij}$
have been renormalized by   $C^{2D}_{ij}$=$L_z$$C^{3D}_{ij}$, in which the $L_z$ is  the length of unit cell along z direction.
The Berry curvatures are   directly calculated from  wave functions  based on Fukui's
method\cite{bm} by using  VASPBERRY code\cite{bm1,bm2}.

\begin{figure*}
  \includegraphics[width=15cm]{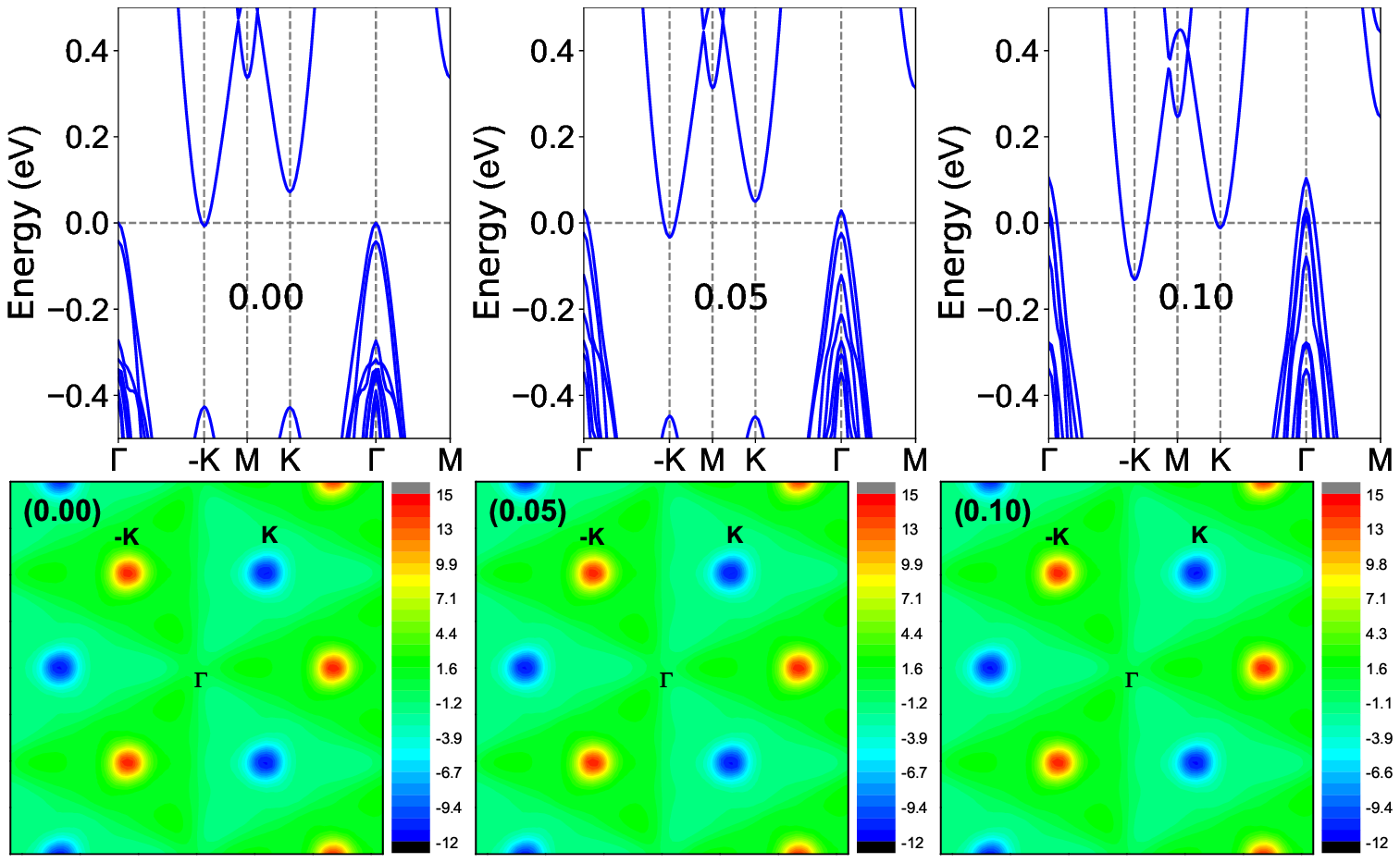}
  \caption{(Color online)For $\mathrm{VGe_2P_4}$ monolayer, the energy band structures (top) and  distribution of Berry curvature in 2D BZ (bottom) at representative $E$=0.00, 0.05 and 0.10 $\mathrm{V/{\AA}}$ using GGA+$U$ ($U$= 3 eV). }\label{band}
\end{figure*}
\section{Structure and stability}
The crystal structures of $\mathrm{VGe_2P_4}$
monolayer are shown in \autoref{st}. Different from experimentally synthesized $\mathrm{MoSi_2N_4}$  with $\alpha_1$ phase\cite{a11}, the $\alpha_2$ structure becomes favorable for $\mathrm{VGe_2P_4}$\cite{a12}. Despite this difference, both $\alpha_2$ and $\alpha_1$ structures lack  inversion symmetry with a space group of $P\bar{6}m2$ (No.187), which allows spontaneous valley polarization.
The $\alpha_2$ can be built by
mirroring $\mathrm{Ge_2P_2}$   double layers of $\alpha_1$ with respect to the vertical surface.
This crystal includes seven covalently bonded atomic layers in the order of
P-Ge-P-V-P-Ge-P along the $z$ axis. In other words,   a $\mathrm{MoS_2}$-like
$\mathrm{VP_2}$ layer is sandwiched between two slightly buckled
honeycomb GeP layers.  The optimized  lattice constants $a$ of $\mathrm{VGe_2P_4}$ is 3.56 $\mathrm{{\AA}}$, which  agrees well with
 with  previous theoretical value\cite{a12}. To obtain the magnetic ground state,  the
total energy difference between  AFM and FM ordering by using rectangle supercell (FIG.1 of ESI)   is calculated, and $\mathrm{VGe_2P_4}$  prefers FM state.

The dynamical stability of the $\mathrm{VGe_2P_4}$ monolayer is proved
by analyzing the phonon spectra, as shown in \autoref{st}.  There are eighteen optical and three acoustical phonon
branches, corresponding to a total of twenty-one branches due to seven
atoms per cell.  It is found that all phonon frequencies are positive in 2D BZ,
confirming the dynamical stability of $\mathrm{VGe_2P_4}$ at
0 K. Therefore, $\mathrm{VGe_2P_4}$ can exist as a free-standing monolayer.
From the application point of view,
the mechanical stability of $\mathrm{VGe_2P_4}$ is checked by elastic constants $C_{ij}$.
Due to  hexagonal symmetry, the  $\mathrm{VGe_2P_4}$ has two independent elastic
constants of $C_{11}$ and $C_{12}$, and the calculated values are $C_{11}$=182.68 $\mathrm{Nm^{-1}}$ and $C_{12}$=51.16 $\mathrm{Nm^{-1}}$. These $C_{ij}$ of $\mathrm{VGe_2P_4}$  satisfy the  Born  criteria of  mechanical stability ($C_{11}>0$ and  $C_{11}-C_{12}>0$)\cite{ela}, confirming  its mechanical stability. The Ab initio molecular
dynamics (AIMD) simulations  are performed to
assess the thermal stability of the $\mathrm{VGe_2P_4}$ monolayer at room
temperature. \autoref{md} shows the free energy
 as a function of the simulation time, along with the snapshots of the geometric structure  at the end of AIMD simulation at 300 K.
The free energy is fluctuated around the equilibrium values without any
sudden changes, and the  small distortions  in the final
configuration are observed.  This confirms the thermodynamical stability of  $\mathrm{VGe_2P_4}$
at room temperature, suggesting the possible room temperature applicability.

\section{magnetic anisotropy and electronic structures}
The magnetization of $\mathrm{VGe_2P_4}$ can affect its symmetry. For out-of-plane FM state,  all
possible vertical mirror symmetry is broken, but the horizontal mirror symmetry is preserved, allowing the spontaneous valley polarization\cite{q14-0,q10-1,q10-2,a6,a7}.
The MAE can be used to decide the  orientation of magnetization of $\mathrm{VGe_2P_4}$. The negative/positive MAE means an easy axis
along the in-plane/out-of-plane direction. The MAE  mainly includes MCA energy $E_{MCA}$  and  MSA energy ($E_{MSA}$).
The SOC can produce a link between the crystalline structure
and the direction of the magnetic moments, giving rise to $E_{MCA}$. Based on GGA+$U$+SOC calculations, the $E_{MCA}$ can be obtained  by $E_{MCA}=E^{||}_{SOC}-E^{\perp}_{SOC}$, where $||$ and $\perp$  mean that spins lie in
the plane and out-of-plane.
 Firstly, a collinear self-consistent  calculation is performed to obtain the convergent charge density without SOC .  Secondly, the convergent charge density is used to carry out  noncollinear non-self-consistent calculation of two different magnetization directions (in-plane and out-of-plane) within SOC.
The $E_{MSA}$ is due to
the anisotropic dipole-dipole (D-D) interaction\cite{a1-7,a7-1}:
 \begin{equation}\label{d-d}
E_{D-D}=\frac{1}{2}\frac{\mu_0}{4\pi}\sum_{i\neq j}\frac{1}{r_{ij}^3}[\vec{M_i}\cdot\vec{M_j}-\frac{3}{r_{ij}^2}(\vec{M_i}\cdot\vec{r_{ij}})(\vec{M_j}\cdot\vec{r_{ij}})]
 \end{equation}
where  the $\vec{M_i}$   represent the local magnetic moments of V atoms, and  vectors $\vec{r_{ij}}$  connects the sites $i$ and $j$.

For a collinear FM monolayer,   the \autoref{d-d} for in-plane case can be expressed as:
 \begin{equation}\label{d-d-1}
E_{D-D}^{||}=\frac{1}{2}\frac{\mu_0M^2}{4\pi}\sum_{i\neq j}\frac{1}{r_{ij}^3}[1-3\cos^2\theta_{ij}]
 \end{equation}
 where $\theta_{ij}$ is the angle between the $\vec{M}$ and $\vec{r_{ij}}$.   For out-of-plane situation, the \autoref{d-d} can further be simplified as:
 \begin{equation}\label{d-d-2}
E_{D-D}^{\perp}=\frac{1}{2}\frac{\mu_0M^2}{4\pi}\sum_{i\neq j}\frac{1}{r_{ij}^3}
 \end{equation}
The $E_{MSA}$  ($E_{D-D}^{||}-E_{D-D}^{\perp}$)can be written as :
\begin{equation}\label{d-d-3}
E_{MSA}=-\frac{3}{2}\frac{\mu_0M^2}{4\pi}\sum_{i\neq j}\frac{1}{r_{ij}^3}\cos^2\theta_{ij}
 \end{equation}
It is clearly seen that  the crystal structure and local magnetic moment decide $E_{MSA}$.  Generally,  the MSA tends to make spins directed parallel to the monolayer.

For V-based $\mathrm{MA_2Z_4}$, $E_{MSA}$ has important effects on  their magnetization direction\cite{gsd,a7-1-1,a7-1-2}.
For $\mathrm{VSi_2P_4}$ monolayer,  an out-of-plane ferromagnet is predicted,  when  only $E_{MCA}$ is considered\cite{q14}. However, the  $\mathrm{VSi_2P_4}$  will prefer in-plane case with including $E_{MSA}$\cite{gsd}. Calculated results show that $\mathrm{VSi_2P_4}$ is a FV semiconductor for out-of-plane case using GGA+$U$ ($U$= 3 eV), and it will become a common magnetic semiconductor for in-plane situation\cite{gsd,q14}. So, the $E_{MSA}$ is considered to decide magnetization direction of $\mathrm{VGe_2P_4}$.
Calculated results show that the MAE is 19 $\mathrm{\mu eV}$ with $E_{MCA}$ and $E_{MSA}$ being 31 $\mathrm{\mu eV}$ and -12 $\mathrm{\mu eV}$, and the  positive  MAE means
that the easy magnetization axis of $\mathrm{VGe_2P_4}$ is out-of-plane.

The spin-polarized band structures of monolayer  $\mathrm{VGe_2P_4}$ by using both GGA+$U$ and GGA+$U$+SOC are calculated. In a trigonal prismatic crystal field environment,  the  V-$d$ orbitals split
into  $d_{z^2}$ orbital, $d_{xy}$+$d_{x^2-y^2}$  and
$d_{xz}$+$d_{yz}$ orbitals, and these orbital characters energy band structures without SOC and with SOC are plotted in \autoref{band0}.
Based on \autoref{band0} (a) and (b),  a distinct
spin splitting can be observed due to the exchange
interaction.  It is found that $\mathrm{VGe_2P_4}$ is  an indirect band
gap semiconductor of 0.091 eV, and its  VBM and CBM  are provided by the spin-dn and spin-up.
The energies of  -K and K  valleys for both conduction and valence bands are degenerate, and no  spontaneous valley polarization is observed.

\begin{figure}
  \includegraphics[width=8cm]{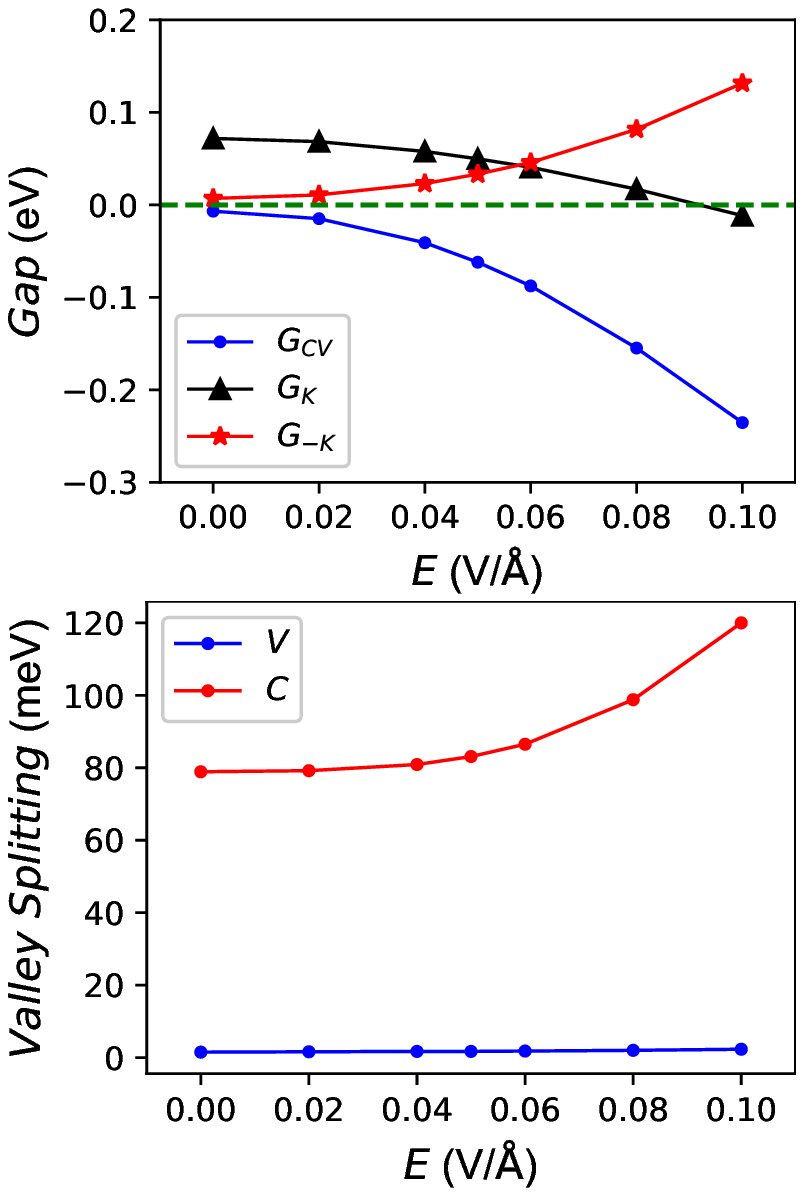}
  \caption{(Color online)For $\mathrm{VGe_2P_4}$ monolayer,  the related  band gap  (top panel)  and  the valley splitting  for both valence and condition bands (bottom panel) as a function of   $E$. }\label{gap}
\end{figure}

The spontaneous  valley polarization is induced by SOC.  Calculated results show observable valley splitting between -K and K valleys  in the conduction bands,  and neglectful valley splitting in the valence bands. This difference can be explained by the distribution of V-$d$ orbitals.
According to \autoref{band0}, it is found that $d_{x^2-y^2}$+$d_{xy}$ orbitals  dominate -K and K valleys of conduction bands, while the ones of valence bands are mainly from $d_{z^2}$ orbitals.
 A simple perturbation theory can be used to further elucidate the
essence of the valley polarization with  SOC Hamiltonian of out-of-plane magnetization  $\hat{H}^0_{SOC}$\cite{q18,v2}:
\begin{equation}\label{m1}
\hat{H}^0_{SOC}=\alpha \hat{L}_z
\end{equation}
in which  $\hat{L}_z$  and  $\alpha$  are the orbital angular moment along $z$ direction and coupling strength.
In light of the orbitals' contribution to -K and K valleys and their wave vector symmetry, the basis functions are chosen as:
\begin{equation}\label{m2}
|\phi^\tau_c>=\sqrt{\frac{1}{2}}(|d_{x^2-y^2}>+i\tau|d_{xy}>)
\end{equation}
\begin{equation}\label{m2}
|\phi^\tau_\nu>=|d_{z^2}>
\end{equation}
where the subscript $c$, $\nu$ and $\tau$ represent  conduction bands, valence bands and  valley index ($\tau=\pm1$).
 The energy levels at -K and K  valleys are then defined as:
\begin{equation}\label{m3}
E^\tau=<\phi^\tau|\hat{H}^0_{SOC}|\phi^\tau>
\end{equation}
 According to the distribution of V-$d$ orbitals, the
valley splitting in the bottom conduction and top valence
bands are given by\cite{q18,v2}:
\begin{equation}\label{m4}
|\Delta E^c|=E^{K}_c-E^{-K}_c=4\alpha
\end{equation}
\begin{equation}\label{m4}
|\Delta E^{\nu}|=E^{K}_{\nu}-E^{-K}_{\nu}=0
\end{equation}
 The perturbation theory results are in
excellent consistency with the first-principle calculations.

According to \autoref{band0} (c), the $\mathrm{VGe_2P_4}$ becomes a metal with a negative  gap ($G_{cv}$) of -7 meV, when the SOC is included.  The Fermi level slightly touches the CBM (-K valley)   and VBM ($\Gamma$ valley), but has no contact with K valley. Therefore, $\mathrm{VGe_2P_4}$ monolayer is a QHVM.
To  describe properties of QHVM, two other  gaps are defined. One is the Fermi level minus the energy of -K valley in the conduction bands ($G_{-K}$), which can describe  concentration of  electron carriers. Another is the energy of K valley in the conduction bands minus the Fermi level ($G_{K}$), which describes how easy the carriers of -K valley jump to K valley. The $G_{-K}$ plus $G_{K}$ equals to valley splitting in the conduction bands. For  $\mathrm{VGe_2P_4}$, the $G_{-K}$ is only about 7 meV, so it needs to be regulated through the external field, for example electric field.

\section{electric field effects}
Electric field can
induce a semiconductor to metal transition  in bilayer $\mathrm{MoSi_2N_4}$ and $\mathrm{WSi_2N_4}$\cite{apl}. Moreover, the increasing electric field  can result in a transition of MA from in-plane to out-of-plane in  $\mathrm{VSi_2P_4}$\cite{jmcc}.  In view of these facts, electric field  may tune physical properties of QHVM in $\mathrm{VGe_2P_4}$. Firstly, we confirm the magnetic ground state under the electric field by the energy differences (per formula unit) between AFM and FM ordering, which are  plotted in FIG.2 of ESI. Within  considered $E$ range,  the FM ordering is always  ground state.
The MAE along with $E_{MCA}$  and $E_{MSA}$   as a function of $E$ are plotted in \autoref{mae}.
It is found that the $E_{MCA}$ is positive within considered $E$ range, which firstly increases, then decreases, and then increases with increasing $E$.
The $E_{MSA}$ can be obtained by \autoref{d-d-3}.  FIG.3 of ESI shows the local magnetic moment of V atom ($M_V$) as a function of $E$. It is found that the $E$ has very little effects on $M_V$ within considered $E$ range, and the change is  only  0.021 $\mu_B$, which leads to  the small change of  $E_{MSA}$  from -11.58 $\mathrm{\mu eV}$ to -12.03 $\mathrm{\mu eV}$. Finally, we calculate the MAE by  $E_{MAE}$=$E_{MCA}$+$E_{MSA}$. It is clearly seen that  $E_{MAE}$  and  $E_{MCA}$ have the same trend with respect to $E$. In considered $E$ range, $E_{MAE}$ is always positive, implying that $\mathrm{VGe_2P_4}$ possesses out-of-plane MA. This confirms spontaneous
 valley polarization in $\mathrm{VGe_2P_4}$ within considered $E$ range.

 The energy band structures and  Berry curvature distribution of $\mathrm{VGe_2P_4}$ under electric field  are investigated, and some representative ones are plotted in \autoref{band}.  The evolutions of related gap ($G_{cv}$, $G_{-K}$ and $G_{K}$)  and  valley splitting for both valence and condition bands  as a function of $E$ are shown in \autoref{gap}.  With increasing $E$,  the $G_{-K}$ increases, which means that  concentration of  electron carriers increases.
 However, the $G_{K}$ decreases, which means that the carriers of -K valley  more easily jump to K valley. When  $E$$>$0.09 $\mathrm{V/{\AA}}$, the Fermi energy level simultaneously touches the -K and K valleys, and the QHVM will disappear. It is found that the increasing $E$ can enhance valley splitting of conduction bands.
 So, the electric field can effectively  tune physical properties of QHVM in  $\mathrm{VGe_2P_4}$.
 By applying an in-plane longitudinal  $E$ in  $\mathrm{VGe_2P_4}$, anomalous velocity $\upsilon$  of Bloch electrons at -K valley  is related with Berry curvature $\Omega(k)$ (see \autoref{band}):$\upsilon\sim E\times\Omega(k)$\cite{xd,qqq}.  The  Berry curvature of -K valley forces
the electron carriers to accumulate on one side of the sample, and  the hole carriers of $\Gamma$ valley move in a straight line.
   When an  out-of-plane  electric fields is applied, the carrier concentration can be effectively tuned.
\begin{figure}
  \includegraphics[width=8cm]{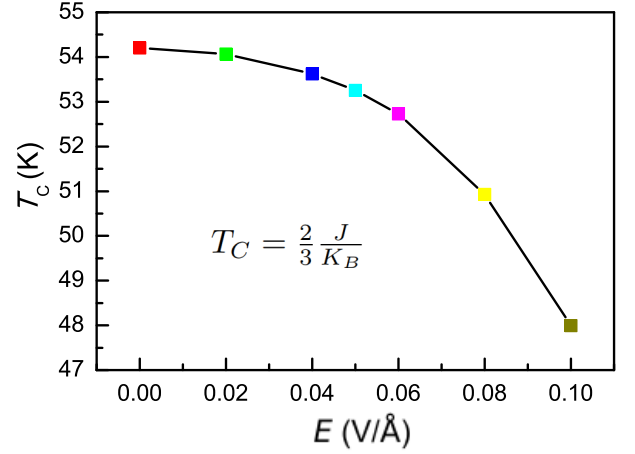}
  \caption{(Color online)For $\mathrm{VGe_2P_4}$ monolayer,  the Curie temperature ($T_C$)  as a function of   $E$. }\label{tc}
\end{figure}

To estimate Curie temperature $T_C$ of $\mathrm{VGe_2P_4}$, it  can be simply
considered as a Ising spin system.  The  transition temperature can be expressed as  $T_C=\frac{2}{3}\frac{J}{K_B}$ by using the mean-field approximations (MFA)\cite{re8}, where $J$ and $K_B$ are  the nearest-neighboring exchange parameter and Boltzmann
constant, respectively.   The $J$ can be obtained from the energy
difference between  AFM  ($E_{AFM}$) and FM ($E_{FM}$) orderings.
According to the FM
and AFM configurations, the  $E_{FM}$ and $E_{AFM}$    can be
written as:
\begin{equation}\label{pe0-1-2}
E_{FM}=E_0-6JS^2
 \end{equation}
  \begin{equation}\label{pe0-1-3}
E_{AFM}=E_0+2JS^2
 \end{equation}
in which $E_0$ is the total energy of systems without magnetic coupling.
The  corresponding $J$ can be attained:
  \begin{equation}\label{pe0-1-3}
J=\frac{E_{AFM}-E_{FM}}{8S^2}
 \end{equation}
The $T_C$ vs $E$ is plotted in \autoref{tc}. Within $E$ range, the predicted $T_C$ is larger than 48 K.

\begin{figure}
  \includegraphics[width=8cm]{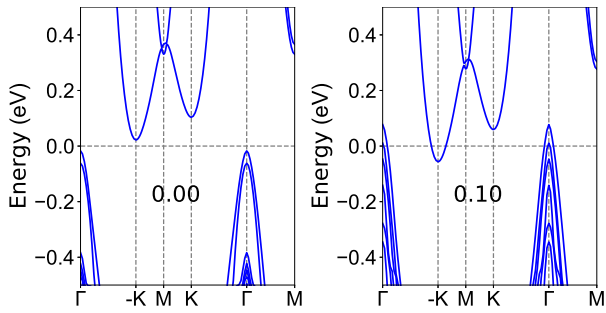}
  \caption{(Color online)For $\mathrm{VGe_2P_4}$ monolayer, the energy band structures  at representative $E$=0.00 and 0.10 $\mathrm{V/{\AA}}$ using GGA+$U$ ($U$= 4 eV). }\label{band-4}
\end{figure}

\section{Discussion and Conclusion}
 For a given material, the correlation strength $U$ should
be determined from future experiment. Here, the $U$= 4 eV is also used  to investigate related properties of $\mathrm{VGe_2P_4}$ to confirm that the QHVM can indeed be achieved by electric field tuning. FIG.4 of ESI shows that  the FM ordering is  ground state, when $E$$<$0.19 $\mathrm{V/{\AA}}$.
The $E_{MAE}$,  $E_{MCA}$  and $E_{MSA}$ along with  the local magnetic moment of V atom ($M_V$)  as a function of $E$ are plotted in FIG.5 and FIG.6 of ESI.
 When $E$$<$0.19 $\mathrm{V/{\AA}}$,  the positive $E_{MAE}$ implies that $\mathrm{VGe_2P_4}$ has out-of-plane MA. The evolutions of related gap ($G_{cv}$, $G_{-K}$ and $G_{K}$)   as a function of $E$ are shown in FIG.7 of ESI.  The representative energy band structures at  $E$$=$0.00 and 0.10 $\mathrm{V/{\AA}}$ are plotted in \autoref{band-4}. Without electric field,  $\mathrm{VGe_2P_4}$ is a FV semiconductor with an indirect band
gap of 0.040 eV. The increasing $E$ can induce FV semiconductor to QHVM transition with critical $E$ of about 0.05 $\mathrm{V/{\AA}}$. However, when $E$$>$0.15 $\mathrm{V/{\AA}}$, the QHVM will disappear. For example $E$=0.10 $\mathrm{V/{\AA}}$, the  $\mathrm{VGe_2P_4}$ is a QHVM with $G_{cv}$, $G_{-K}$ and $G_{K}$ being
-0.133 eV, 0.056 eV and 0.060 eV. When correlation strength $U$ is in the reasonable range of 3 eV to 4 eV, the  $\mathrm{VGe_2P_4}$ can be tuned into QHVM by electric field.

In summary, we have proposed the concept of QHVM, which may be realized in monolayer  $\mathrm{VGe_2P_4}$ with dynamical, mechanical and thermal  stabilities.
The electric field can be used tune  the related physical properties of QHVM in  $\mathrm{VGe_2P_4}$.
 For a certain electric field region, QHVM properties   can be maintained, and the carrier concentration of -K valley can be effectively tuned.  When  in-plane and out-of-plane  electric fields are applied,  the  electron carriers  of -K valley  transversely drift and eventually accumulate on one edge of the sample, while the hole carriers of $\Gamma$ valley longitudinally move along the in-plane electric field direction.
 Our findings can inspire more works to search for QHVM.

~~~~\\
~~~~\\
\textbf{Conflicts of interest}
\\
There are no conflicts to declare.

\begin{acknowledgments}
This work is supported by Natural Science Basis Research Plan in Shaanxi Province of China  (2021JM-456). We are grateful to Shanxi Supercomputing Center of China, and the calculations were performed on TianHe-2.
\end{acknowledgments}


\begin{references}

 \bibitem{b0}J. R. Schaibley, H. Yu, G. Clark, P. Rivera, J. S. Ross, K. L. Seyler,
W. Yao and X. Xu, Nat. Rev. Mater. \textbf{1}, 16055 (2016).


\bibitem{b1}L. J. Sham, S. J. Allen, A. Kamgar, and D. C. Tsui, Phys. Rev.
Lett. \textbf{40}, 472 (1978).


\bibitem{b2} S. A. Wolf, D. D. Awschalom, R. A. Buhrman, J. M. Daughton,
S. von Moln$\acute{a}$r, M. L. Roukes, A. Y. Chtchelkanova, and D. M.
Treger, Science \textbf{294}, 1488 (2001).

\bibitem{q8-1}A. Srivastava, M. Sidler, A. V. Allain, D. S. Lembke, A. Kis,
and A. Imamoglu, Nat. Phys. \textbf{11}, 141 (2015).

\bibitem{q8-2} K. F. Mak, K. He, J. Shan, and T. F. Heinz, Nat. Nanotechnol.
\textbf{7}, 494 (2012).


\bibitem{q8-3}H. Zeng, J. Dai, W. Yao, D. Xiao, and X. Cui, Nat. Nanotechnol.
\textbf{7}, 490 (2012).

\bibitem{q9-1}8 H. Zeng, J. Dai, W. Yao, D. Xiao and X. Cui, Nat.
Nanotechnol. \textbf{7}, 490 (2012).


\bibitem{q9-2}M. Zeng, Y. Xiao, J. Liu, K. Yang and L. Fu, Chem. Rev. \textbf{118}, 6236  (2018).



\bibitem{q9-3} C. Zhao, T. Norden, P. Zhang, P. Zhao, Y. Cheng, F. Sun,
J. P. Parry, P. Taheri, J. Wang, Y. Yang, T. Scrace, K. Kang,
S. Yang, G. Miao, R. Sabirianov, G. Kioseoglou, W. Huang,
A. Petrou and H. Zeng, Nat. Nanotechnol.  \textbf{12}, 757 (2017).


\bibitem{q9-4}D. MacNeill, C. Heikes, K. F. Mak, Z. Anderson,
A. Korm$\acute{a}$nyos, V. Z$\acute{o}$lyomi, J. Park and D. C. Ralph, Phys.
Rev. Lett. \textbf{114}, 037401 (2015).


\bibitem{q10}W. Y. Tong, S. J. Gong, X. Wan, and C. G. Duan,
Nat. Commun. \textbf{7}, 13612 (2016).

\bibitem{q11}Y. B. Liu, T. Zhang, K. Y. Dou, W. H. Du, R. Peng, Y. Dai, B. B. Huang,
and Y. D. Ma, J. Phys. Chem. Lett. \textbf{12}, 8341 (2021).

\bibitem{q12}Z. Song, X. Sun, J. Zheng, F. Pan, Y. Hou, M.-H. Yung, J. Yang,
and J. Lu, Nanoscale \textbf{10}, 13986 (2018).


\bibitem{q13}J. Zhou, Y. P. Feng, and L. Shen, Phys. Rev. B \textbf{102}, 180407(R)
(2020).


\bibitem{q13-1}P. Zhao, Y. Ma, C. Lei, H. Wang, B. Huang, and Y. Dai,
Appl. Phys. Lett. \textbf{115}, 261605 (2019).

\bibitem{q19}S. D. Guo, J. X. Zhu, W. Q. Mu and B. G. Liu, Phys. Rev. B \textbf{104}, 224428 (2021).


\bibitem{gsd}S. D. Guo,  Y. L. Tao,  K. Cheng, B. Wang and Y. S. Ang, J. Phys.: Condens. Matter (2022). https://doi.org/10.1088/1361-648X/ac9c3d

\bibitem{q14}X. Y. Feng, X. L. Xu, Z. L. He, R. Peng, Y. Dai, B. B. Huang and Y. D. Ma,  Phys. Rev. B \textbf{104}, 075421 (2021).
\bibitem{q14-0}S. Li, Q. Q. Wang, C. M. Zhang, P. Guo and S. A. Yang,  Phys. Rev. B  \textbf{104}, 085149 (2021).

\bibitem{q14-0-1}Q. R. Cui, Y. M. Zhu, J. H. Liang, P. Cui and H. X. Yang,  Phys. Rev. B  \textbf{103}, 085421 (2021).
\bibitem{q14-0-2}Y. L. Wang and  Y. Ding, Appl. Phys. Lett. \textbf{119}, 193101 (2021).


\bibitem{q14-0-3}X. Zhou, R. Zhang, Z. Zhang, W. Feng, Y. Mokrousov and Y. Yao,  npj Comput. Mater. \textbf{7}, 160
(2021).


\bibitem{q14-1}H. X. Cheng, J. Zhou, W. Ji, Y. N. Zhang and Y. P. Feng, Phys. Rev. B \textbf{103}, 125121 (2021).



\bibitem{q17}W. Du, Y. Ma, R. Peng, H. Wang, B. Huang, and Y. Dai,
J. Mater. Chem. C \textbf{8}, 13220 (2020).

\bibitem{q18}R. Li, J. W. Jiang, W. B. Mi  and H. L. Bai, Nanoscale  \textbf{13}, 14807 (2021).

\bibitem{q14-2}K. Sheng, Q. Chen, H. K. Yuan and Z. Y. Wang, Phys. Rev. B \textbf{105}, 075304 (2022).


\bibitem{q14-3}P. Jiang, L. L. Kang,  Y. L. Li,  X. H. Zheng,  Z. Zeng  and S. Sanvito,  Phys. Rev. B \textbf{104}, 035430 (2021).

\bibitem{q14-4}K. Sheng, H. K. Yuan and B. K. Zhang, Nanoscale (2022).DOI: 10.1039/d2nr03860a



\bibitem{q15}Y. Zang, Y. Ma, R. Peng, H. Wang, B. Huang, and Y. Dai,
Nano Res. \textbf{14}, 834 (2021).



\bibitem{q16}R. Peng, Y. Ma, X. Xu, Z. He, B. Huang, and Y. Dai, Phys. Rev.
B \textbf{102}, 035412 (2020).




\bibitem{q10-1} H. Hu, W. Y. Tong, Y. H. Shen, X. Wan, and C. G. Duan, npj
Comput. Mater. \textbf{6}, 129 (2020).

\bibitem{q10-2}K. Sheng, B. K. Zhang, H. K. Yuan and Z. Y. Wang, Phys. Rev. B  \textbf{105}, 195312 (2022).
\bibitem{a6}S. D. Guo, J. X. Zhu, M. Y. Yin and B. G. Liu, Phys. Rev. B  \textbf{105}, 104416 (2022).

\bibitem{a7}S. D. Guo, W. Q. Mu and B. G. Liu, 2D Mater.  \textbf{9}, 035011 (2022).


\bibitem{xd} X. Xu, W. Yao, D. Xiao and T. F. Heinz, Nat. Phys. \textbf{10}, 343 (2014).

\bibitem{qqq}D. Xiao, M. C. Chang, and Q. Niu, Rev. Mod. Phys. \textbf{82}, 1959
(2010).

\bibitem{a11}Y. L. Hong, Z. B.  Liu, L. Wang  T. Y. Zhou,  W. Ma, C. Xu, S. Feng,
L. Chen, M. L. Chen, D. M. Sun, X. Q. Chen, H. M. Cheng and W. C. Ren, Science  \textbf{369}, 670 (2020).

\bibitem{a12}L. Wang, Y. Shi, M. Liu, A. Zhang, Y.-L. Hong, R. Li,
Q. Gao, M. Chen, W. Ren, H.-M. Cheng, Y. Li, and X.-
Q. Chen, Nat. Commun. \textbf{12}, 2361 (2021).






\bibitem{1}P. Hohenberg and W. Kohn, Phys. Rev. \textbf{136},
B864 (1964); W. Kohn and L. J. Sham, Phys. Rev. \textbf{140},
A1133 (1965).



\bibitem{pv1} G. Kresse, J. Non-Cryst. Solids \textbf{193}, 222 (1995).

\bibitem{pv2} G. Kresse and J. Furthm$\ddot{u}$ller, Comput. Mater. Sci. 6, \textbf{15} (1996).

\bibitem{pv3} G. Kresse and D. Joubert, Phys. Rev. B \textbf{59}, 1758 (1999).
\bibitem{pbe}J. P. Perdew, K. Burke and M. Ernzerhof, Phys. Rev. Lett. \textbf{77}, 3865 (1996).


\bibitem{u}S. L. Dudarev, G. A. Botton, S. Y. Savrasov, C. J. Humphreys and A. P. Sutton, Phys. Rev. B \textbf{57}, 1505 (1998).

\bibitem{pv5}A. Togo, F. Oba, and I. Tanaka, Phys. Rev. B \textbf{78}, 134106
(2008).

\bibitem{bm}T. Fukui, Y. Hatsugai and H. Suzuki,  J. Phys. Soc. Japan. \textbf{74},
1674 (2005).


\bibitem{bm1}H. J. Kim,  https://github.com/Infant83/VASPBERRY, (2018).
\bibitem{bm2}H. J. Kim, C. Li, J. Feng, J.-H. Cho, and Z. Zhang, Phys. Rev. B  \textbf{93}, 041404(R) (2016).

\bibitem{ela}E. Cadelano and L. Colombo, Phys. Rev. B  \textbf{85}, 245434 (2012).



\bibitem{a1-7}X. B. Lu, R. X. Fei, L. H. Zhu and  L. Yang, Nat. Commun. \textbf{11}, 4724 (2020).




\bibitem{a7-1}K. Yang, G. Y. Wang, L. Liu ,  D. Lu  and H. Wu, Phys. Rev. B  \textbf{104}, 144416 (2021).


\bibitem{a7-1-1}S. D. Guo, W.-Q. Mu, J.-H. Wang, Y.-X. Yang, B. Wang and Y.-S. Ang, Phys. Rev. B  \textbf{106}, 064416 (2022).


\bibitem{a7-1-2}D. Dey, A. Ray and L. P. Yu, Phys. Rev. Materials \textbf{6}, L061002 (2022).




\bibitem{v2}P. Zhao, Y. Dai, H. Wang, B. B. Huang and  Y. D. Ma, ChemPhysMater, \textbf{1}, 56 (2022).


\bibitem{apl}Q. Y. Wu,  L. M. Cao,  Y. S. Ang and  L. K. Ang, Appl. Phys. Lett. \textbf{118}, 113102 (2021).

\bibitem{jmcc}S. D. Guo,   X. S. Guo,   G. Z. Wang,   K. Cheng  and  Y. S. Ang, J. Mater. Chem. C (2022). https://doi.org/10.1039/D2TC03293G

\bibitem{re8}L. Ke, B. N. Harmon, and M. J. Kramer, Physical Review B \textbf{95}, 104427 (2017).


\end{references}
\end{document}